\documentclass[prb,aps,superscriptaddress,twocolumn]{revtex4-2}
\usepackage{float}
\usepackage{csquotes}
\usepackage{wrapfig}
\usepackage{xfrac}
\usepackage{graphicx}[h]
\usepackage{xcolor}
\usepackage[normalem]{ulem}
\usepackage{graphicx}
\usepackage{dcolumn}
\usepackage{bm}
\usepackage{subfigure}

\begin{document}

\title{Efficient method to calculate energy spectra for analysing magneto-oscillations}
\author{Hamed Gramizadeh}
\affiliation{Department of Engineering, Reykjavik University, Menntavegi 1, IS-102 Reykjavik, Iceland}
\author{Denis R. Candido}
\affiliation{Department of Physics and Astronomy, University of Iowa, Iowa City, Iowa 52242, USA}
 \author{Andrei Manolescu}
 \affiliation{Department of Engineering, Reykjavik University, Menntavegi 1, IS-102 Reykjavik, Iceland}
 \author{J.\ Carlos Egues}
 \affiliation{Instituto de F\'isica de S\~ao Carlos, Universidade de S\~ao Paulo, 13560-970
S\~ao Carlos, SP, Brazil}
\affiliation{Department of Physics, University of Basel, CH-4056, Basel, Switzerland}
\author{Sigurdur I.\ Erlingsson}%
\affiliation{Department of Engineering, Reykjavik University, Menntavegi 1, IS-102 Reykjavik, Iceland}
 \email{Sie@ru.is}
\date{\today}%
\begin{abstract}
Magneto-oscillations in two-dimensional systems with spin-orbit interaction are typically characterized by fast Shubnikov-de~Haas (SdH) oscillations and slower spin-orbit-related beatings.  The characterization of the full SdH oscillatory behavior in systems with both spin-orbit interaction and Zeeman coupling requires a time consuming diagonalization of large matrices for many magnetic field values.
By using the Poisson summation formula we can explicitly separate the density of states into, fast and slow oscillations, which determine the corresponding fast and slow parts of the magneto-oscillations.  We introduce an efficient scheme of partial diagonalization of our Hamiltonian, where only states close to the Fermi energy are needed to obtain the SdH oscillations, thus reducing the required computational time.   This allows an efficient method for fitting numerically the SdH data, using the inherent separation of the fast and slow oscillations.  
We compare systems with {\em only} Rashba spin-orbit interaction (SOI) and {\em both} Rashba and Dresselhaus SOI with, and without, an in-plane magnetic field.  The energy spectra are characterized in terms of symmetries, which have direct and visible consequences  in the magneto-oscillations.
To highlight the benefits of our methodology, we use it to extract the spin-orbit parameters by fitting realistic transport data.  
\end{abstract}

\maketitle

\section{Introduction}
\label{sec:level1}
Shubnikov-de~Haas (SdH) oscillations \cite{sdh-original,sdh-original2} have been an important tool to characterize charge densities, and scattering times in 2D semiconductor \cite{ihn10:book}.  In addition, the SdH oscillations have  been used to extract the Rashba and Dresselhaus spin-orbit interactions (SOI)\cite{winkler03:1}.
Earlier theoretical description showed that the SOI leads to changes in the oscillation beating pattern \cite{das1989evidence}, and further analysis of the same group incorporated the known exact result\cite{bychkov1984oscillatory} to improve the analysis of the Rashba and Zeeman coupling \cite{das90:8278}.
As is pointed out in Ref.~\cite{das1990zero}, the study and interpretation of oscillations in the magnetoresistance relies on  some assumptions, as for example, what the dominant source of SOI is. 
A method that has often been used to estimate the strength of the Rashba coupling was introduced in Ref.\ \cite{nitta97:1335,engels1997experimental,schaepers98:4324}, which uses the density of states (DOS) at {\em zero} magnetic field to relate the DOS to the Rashba SOI strength $\alpha$. However, this method has drawbacks since it can not account for Zeeman (via the g-factor $g^*$) or Dresselhaus spin-orbit coupling \cite{dresselhaus1955spin,gilbertson2008zero}. There have been some attempts to analyze the SdH oscillations in terms of $\alpha$, $\beta$, and $g^*$ , but they have mostly involved qualitative comparison with the energy spectrum of pure Rashba and pure Dresselhaus \cite{gilbertson2008zero,akabori2006spin}.

Magnetoresistance oscillations were considered by Tarasenko and co-authors~\cite{
averkiev05:543,tarasenko02:552} for the special case of $\alpha$ = $\beta$ and no Zeeman coupling. They showed that the beatings vanished for this case, since the corresponding spectrum consists of equally spaced Landau levels. Furthermore, the effects of Zeeman splitting and tilted magnetic field (in the absence of spin-orbit coupling) were considered in Ref.\ \cite{tarasenko02:1769}.
In Ref.~\onlinecite{yang06:045303}, full numerical calculations of  magneto-oscillations were performed  for relatively high magnetic fields and low electron densities, which is far away from the regime of recent experimental works.\cite{beukmann17:241401}
In Ref.~\cite{beukmann17:241401,fal1992cyclotron} numerical calculations of magnetoresistivity-oscillations were performed, but a general analysis of the oscillations, relating the frequency and position of the beating pattern directly to $\alpha$ and $\beta$, was not presented. Such connections are very important for experimental works as they allow the extraction of system parameters. 
In a recent experimental work, SdH oscillations were considered in InAs 2DEGs, where the Rashba SOI was tuned, but there were unresolved issues concerning the cubic Dresselhaus SOI \cite{beukmann17:241401}.  Furthermore,  the effects of the tilted magnetic field were theoretically considered in the context of the cyclotron and electric-dipole spin resonances in the presence of both Rashba and Dresselhaus SOI~\cite{fal1992cyclotron}. For tilting angles at which the Zeeman splitting and cyclotron energy were equal, the effects of the SOI could be made more pronounced.  This has been used in more recent experiments studying magnetization~\cite{herzog17:103012} and magneto-oscillations~\cite{wilde2013alternative}, although  the analysis suffers from the same issues  discussed in Ref.\ \cite{gilbertson2008zero}. 

In this paper we  introduce a new efficient method to obtain the relevant energy spectrum for magneto transport, in the presence of both Rashba and Dresselhaus SOIs and Zeeman coupling. Our method is based on the diagonalization of a partial/truncated Hamiltonian, and  allows a faster calculation, and clearer interpretation of SdH magneto-oscillations.  In Sec.\ \ref{sec:spectrum}  we introduce the system properties and the partial Hamiltonian.  In Sec.\ \ref{sec:DOS} we present the density of states using the Poisson summation formula and highlight the fast and slow, oscillations. 
 Finally, we apply our method to accurately fit realistic magneto-oscillation data, highlighting the speed and convenience of our method.

\section{Hamiltonian and numerical diagonalization}
\label{sec:spectrum}
Our focus will be on two dimensional electron gas (2DEG) in the presence of a magnetic field $\bm{B}=( B_\parallel \cos(\phi),B_\parallel \sin(\phi),B_\perp)$, where $B_\perp$ is the component of the magnetic field perpendicular to the 2DEG.  In addition, we  consider both Rashba\cite{bychkov1984oscillatory} and Dresselhaus\cite{dresselhaus1955spin} spin-orbit couplings.  The resulting Hamiltonian is 
\begin{eqnarray}
 H_\mathrm{2D}&=&\frac{1}{2m^*} \left ( \pi_x^2 + \pi_y^2
 \right )+\frac{g^*\mu_B}{2} B_\perp\sigma_z  \nonumber \\
  & +&  \frac{g^*\mu_B B_\parallel}{2} \left ( \sigma_x \cos(\phi) +\sin (\phi) \sigma_y \right )   \nonumber  \\
 &+&\frac{\alpha}{\hbar}(\pi_y \sigma_x-\pi_x \sigma_y)+\frac{\beta}{\hbar}(\pi_x \sigma_x-\pi_y \sigma_y),
 \label{eq:Hfull}
\end{eqnarray}
where the $\hbar$ is reduced Planck's constant, $m^*$ is the effective electron mass, $g^*$ is the effective $g$-factor, and $\mu_B$ is the Bohr magneton, and $\sigma_x$, $\sigma_y$, $\sigma_z$ denote the usual  Pauli matrices.
The angle $\theta$ describes the tilting of the magnetic field away from the perpendicular direction, and we assume that $B \equiv B_\perp$ is fixed for all tilting angles, which is done to ease the comparison between different tilting angles, with absolute value of the applied magnetic field $B/\cos(\theta)$.
The strength of the Rasbha and Dresselhaus SOI are determined by the coefficients $\alpha$ and $\beta$, respectively. The momenta are given by $\pi_x=p_x-eB y/2$, and $\pi_y=p_y+eB x /2$, where $e>0$ is the electrical charge.  Note that the gauge is chosen such that $B_\parallel$ drops out from the momenta once the 3D problem is projected onto the lowest transverse level.  Next, we introduce the ladder operators
\begin{equation}
a=\frac{\ell_c}{\sqrt{2}\hbar} \left  (\pi_x -i \pi_y \right ), \mathrm{\quad and \quad}
a^\dagger=\frac{\ell_c}{\sqrt{2}\hbar} \left  (\pi_x +i \pi_y \right ),
\end{equation} 
where $\ell_c=\sqrt{\frac{\hbar}{eB}}$ is the magnetic length.  The ladder operators obey the commutation relation $[a,a^\dagger]=1$, as a consequence of the canonical commutation relations $[x,p_x]=i\hbar$ and $[y,p_y]=i\hbar$. The Hamiltonian then 
reduces to
\begin{eqnarray}
\frac{H_\mathrm{2D}}{\hbar \omega_c} &=&a^{\dagger}a+\frac{1}{2}+\frac{\tilde{\Delta}}{2} \left ( \sigma_z+\frac{\tan(\theta)}{2} \left (\sigma_+ e^{i\phi}+ \sigma_- e^{-i\phi} \right ) \right ) \nonumber \\
+&& \!\!\!\!\!\!\!\!\! \frac{\beta}{\sqrt{2}\hbar \omega_c \ell_{c}}(a^{\dagger}\sigma_{+}+a\sigma_{-})  - \frac{i\alpha}{\sqrt{2}\hbar \omega_c \ell_{c}}(a^{\dagger}\sigma_{-}-a\sigma_{+}), \nonumber  \\
\label{eq:HfullLadder}
\end{eqnarray}
where the Zeeman term $\widetilde{\Delta}=\frac{g^* \mu_B B}{\hbar \omega_c}=\frac{g^{*}m^*}{2}$ inherited its sign from the $g^{*}$-factor, and $\omega_{c}=eB/m^*$ is
the cyclotron frequency, $\sigma_\pm=\sigma_x \pm i \sigma_y$ . 
\begin{figure}[t]
\centering
\includegraphics[angle=-0,width=0.49\textwidth]{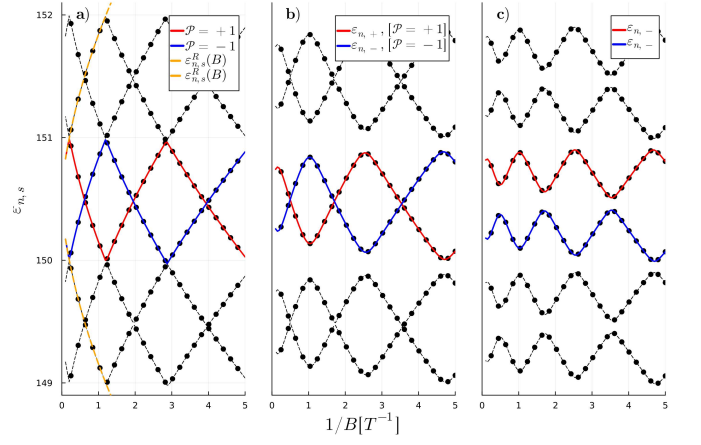}
\caption{Comparison of full diagonalization [black points] and partial Hamiltonian [red and blue curves], for a) $\alpha=7.5$\,meV\,nm, $\beta=0$, b) $\alpha=7.5$\,meV\,nm, $\beta=3.0$\,meV\,nm, and c) $\alpha=7.5$\,meV\,nm, $\beta=3.0$\,meV\,nm, and $\theta=\pi/3$. Other parameters are $m^*=0.04$, $g^*=-12$ and $n_\mathrm{2D}=0.0176$\,nm$^{-2}$ for InAs based systems.\cite{beukmann17:241401}}
\label{fig:Fig1}
\end{figure}
The standard way of obtaining the spectrum of the Hamiltonian  Eq.\ (\ref{eq:HfullLadder}) is by creating a  matrix of dimension $2N \times 2N$, where $N$ is the number of eigenstates of $a^\dagger a $ (i.e.\ $a^\dagger a |m \rangle =m |m \rangle $, $m=0, 1, \dots, N-1)$, in addition to accounting for the spin-degree (i.e\ $\sigma_z |\sigma \rangle =\sigma|\sigma\rangle $, $\sigma=\pm 1$).  The choice of $N$ depends on the number of eigenstates that are required for a given problem.  In the case of magnetotransport calculations for realistic systems parameters, the required eigenstates are counted in the hundreds, and to calculate those states accurately, the size of $N$ should be around four times larger \cite{golub:book}, resulting in $N \sim 10^3$.  Although diagonalizing a single such matrix does not represent a computational challenge, the diagonalization has to be repeated for multiple values of magnetic field (measured in the thousands), and $\alpha$, $\beta$, etc.  Accounting for all this, calculating a set of magnetoresistance curves can lead to computational time around multiple hours \footnote{Here we benchmark using a powerful laptop.}. 

The method we introduce here is designed to efficiently calculate the eigenenergies for a given $n$, which labels the Landau levels.  Before outlining the methods, we first discuss general properties of the Hamiltonian  Eq. (\ref{eq:HfullLadder}).   If we have $\beta=\theta=0$, we can obtain exact eigenvalues (see App.\ \ref{app:Rashba})
\begin{eqnarray}
\varepsilon _{n,+}&=&n+1 - \sqrt{\frac{(1-\widetilde{\Delta})^2}{4} + 4\frac{\varepsilon_R}{\hbar \omega_c } (n+1)}, \label{eq:Enp}\\
\varepsilon _{n,-}&=&n +\sqrt{\frac{(1-\widetilde{\Delta})^2}{4} + 4\frac{\varepsilon_R}{\hbar \omega_c } n}, \label{eq:Enm}
\end{eqnarray}
where $\varepsilon_R=\frac{m^*\alpha^2}{2\hbar^2}$. These eigenvalues are plotted in Fig.\ \ref{fig:Fig1}a) for $n=150$ (dashed orange curve).  When the same system is diagonalized numerically, the energy spectra take a sawtooth shape since the numerical diagonalization orders the eigenvalues according to their size and crossings turn into anticrossings (black dotted lines).  There is an underlying parity symmetry for $\theta=0$, first introduced in Refs.\ \citep{casanova2010deep,braak11:100401} for $\alpha=\beta$, and later extended for systems with Rashba and Dresselhaus coupling in Ref.\ \cite{candido23:cond-mat}. This parity allows the spectrum to  be split into two separate subspaces that can be diagonalized separately, see App.\ \ref{app:HPD}.  When this is done, we obtain states with different parities crossing each other, as they belong to different parity subspaces (blue and red curves). However,  they anticross with other states that belong to the {\em same} parity space.

In Fig.\ \ref{fig:Fig1}b) a non-zero value of $\beta=3.0$\,meV\,nm is added, which opens up overall gaps in the spectrum, but leaves some crossing unaffected.  The spectrum now consists of pairs of states for each value of $n$ and $s=\pm 1$ which cross, but anticross with adjacent states above and below.
\begin{figure}[htb!]
\includegraphics[angle=0,width=0.4\textwidth]{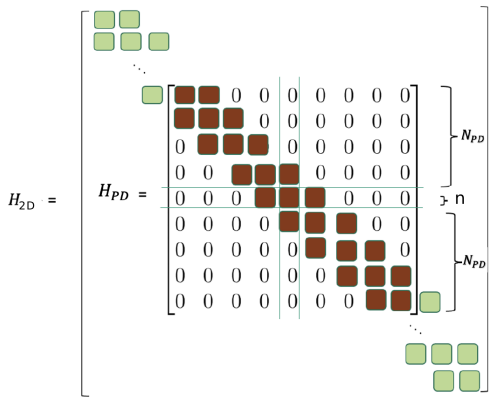}
\caption{The structure of $H_\mathrm{PD}$ illustrated relative to the full matrix $H_\mathrm{2D}$.  For a given value of $n$ the partial matrix $H_\mathrm{PD}$ is constructed around matrix element $[H_\mathrm{2D}]_{n,n}$.}
\label{fig:Fig2}
\end{figure}
Finally, in Fig.\ \ref{fig:Fig1}c) an in-plane component of the magnetic field is added with $\theta=\pi/3$.  For this case, the parity is no longer a good quantum number, i.e.\ the parity operator does not commute with $H_{2D}$, and extra anticrossings opens up between  $|n,+ \rangle$ and $|n,- \rangle$ states corresponding to eigenenergies $\varepsilon_{n,+}$ and $\varepsilon_{n,+}$, respectively.

\subsection{Numerical methods}
Now, we turn to describing the numerical diagonalization procedure.  As can be seen in Fig.\ \ref{fig:Fig1}b) and c) the eigenstates are always pushed up or down by their couplings to adjacent  states.  This results in each state following a unique curve which can be tracked, as a function of $1/B$, for all $n$.  Based on this, we introduce a partial diagonalization outlined in Fig.\ \ref{fig:Fig2}, where $n=0,1,2 \dots $ is the Landau level index of interest.  The matrix representation of Eq.\ (\ref{eq:HfullLadder}) can be written as a block-tridiagonal matrix with diagonal $2 \times 2$ blocks
\begin{eqnarray}
    [H_\mathrm{2D}]_{m,m}
    &=&  \left [
    \begin{array}{cc}
     m +\frac{1}{2}+ \frac{\tilde{\Delta}}{2} & \frac{\tilde{\Delta}}{2}\tan(\theta)e^{i \phi}  \\
     \frac{\tilde{\Delta}}{2}\tan(\theta)e^{-i \phi}  & m +\frac{1}{2} - \frac{\tilde{\Delta}}{2} 
    \end{array}
    \right ],
    \label{eq:H2Dmm}
\end{eqnarray}
where $\sigma_0$ is the Pauli identity matrix, and off-diagonal $2\times 2$ block is given by
\begin{eqnarray}
    [H_\mathrm{2D}]_{m,m+1}
    &=&\sqrt{m+1} \frac{1}{\sqrt{2}\hbar \omega_c \ell_c}
    \left [
    \begin{array}{cc}
     0 & 2\beta \\
     -2i \alpha & 0
    \end{array}
    \right ].
    \label{eq:H2Dmm1}
\end{eqnarray}
With these we construct the partial matrix $H_\mathrm{PD}$ centered on block $n$ with $N_\mathrm{PD}$ blocks above and below.  The resulting matrix has dimension $2(2N_\mathrm{PD}+1) \times 2(2N_\mathrm{PD}+1)$.

\begin{figure}[t]
\centering
\includegraphics[angle=-0,width=0.48\textwidth]{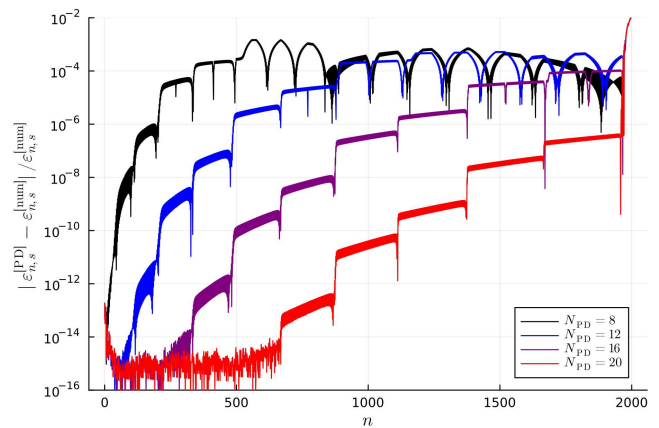}
\caption{Relative deviation between eigenstates obtained using full numerical diagonalization with $N=1000$, $\varepsilon_{n,s}^{[\mathrm{num}]}$ and the eigenstates with the partial diagonalization, $\varepsilon_{n,s}^{[\mathrm{PD}]}$, as a function of $N_\mathrm{PD}$ for magnetic field $B=0.15$\,T.  Parameter values are $\alpha=7.5$\,meV\,nm, $\beta=3.0$\,meV\,nm, and $\theta=0$. Other parameters are $m^*=0.04$, $g^*=-12$ and $n_\mathrm{2D}=0.0176$\,nm$^{-2}$ for InAs based. systems\cite{beukmann17:241401}}
\label{fig:Fig3}
\end{figure}

If the parity is a good quantum number, i.e.\ $\theta=0$, then each block in $H_\mathrm{PD}$ is halved (i.e.\ becomes $1 \times 1$) when each parity subspace is considered, see App.\ \ref{app:HPD} for details. For states with $n \leq N_\mathrm{PD}$, then the lower part of the partial matrix is decreased accordingly, and for $n=0$ only $N_\mathrm{NP}$ states above $n$ are needed.  With this, the entire spectrum can be calculated for each value of $n$.
To test the accuracy of this procedure we calculate the relative deviation between the full numerical diagonalization, $\varepsilon_{n,s}^{[\mathrm{num}]}$ for $N=1000$ and the eigenstates obtained with the partial diagonalization, $\varepsilon_{n,s}^{[\mathrm{PD}]}$, at $B=0.15$\,T for $\alpha=7.5$\,meV\,nm, $\beta=3.0$\,meV\,nm, and $\theta=0$. Figure \ref{fig:Fig3} shows our results for $N_{\mathrm{PD}}=8$, $12$, $16$ and $20$.
Already for $N_\mathrm{PD}=16$ the relevant eigenenergies (first quarter of eigenvalues) have a relative deviation less than $10^{-10}$, and for $N_\mathrm{PD}=20$ the machine precision is reached for all relevant eigenvalues.

As we will see in the next section, allowing $n$ to take   {\em non-integer} values can be useful in calculating the density of states and transport properties.  As is discussed in App.\ \ref{app:HPD} this can be implemented via the partial diagonalization, i.e.\ one can calculate eigenenergies $\varepsilon_{n+\Delta x,s}$, where $\Delta x \in [-0.5,0.5]$ is a real number.  The interval is set by the condition that $\varepsilon_{n+\Delta x,s}=\varepsilon_{n+1-\Delta x,s}$, i.e.\  $\Delta x=0.5$ corresponds to a crossing with the next state above, and similarly $\Delta x=-0.5$ corresponds to a crossing with the next state below.

\section{Density of States and $F$-function}
\label{sec:DOS}
\begin{figure}[hbt!]
\centering
\includegraphics[angle=-0,width=0.49\textwidth]{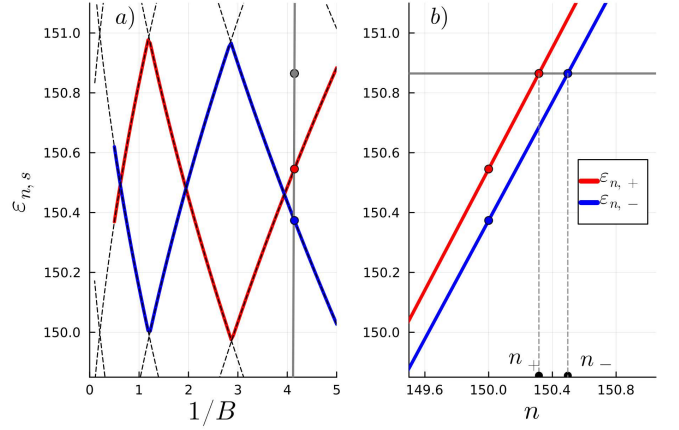}
\caption{a) Energy level $n=150$ using the PD algorithm, along with adjacent states [dashed black line] for $\alpha=7.5$\,meV\,nm and $\beta=0.0$.  The solid gray line shows $E_F/\hbar \omega_c$, and its value at $1/B=4.145$ [gray circle]. b) The energy levels $\varepsilon_{n,s}(B)$ as a function of $n$ showing the intersection with $E_F/\hbar \omega_c$ at $1/B=4.145$. Other parameters are $m^*=0.04$, $g^*=-12$ and $n_\mathrm{2D}=0.0176$\,nm$^{-2}$ for InAs based systems.\cite{beukmann17:241401}}
\label{fig:Fig4}
\end{figure}

The eigenenergies of the Hamiltonian  Eq.\ (\ref{eq:Hfull}) results in a discrete spectrum, the well known Landau levels \cite{ihn10:book}.   The resulting DOS is given by
\begin{eqnarray}
 D(E_F,B)&=& \frac{1}{2\pi \ell_c^2}\sum_{n=0}^{\infty}\sum_s L_\Gamma(E_F-\hbar \omega_c\varepsilon_{ns}(B)) , 
\label{eq:DOSdef}
\end{eqnarray}
where $1/2 \pi \ell_c^2$ accounts for the Landau level degeneracy (per spin), and $L_\Gamma(x)$ describes broadening due to impurity scattering \cite{ihn10:book}. Here it is assumed that all levels are broadened by a phenomenological parameter $\Gamma$, e.g., $\delta(  \cdot) \rightarrow L_\Gamma(\cdot)$, for Gaussian broadening with $L_\Gamma(x)=e^{-\frac{x^2}{2 \Gamma^2}}/\sqrt{2 \pi \Gamma^2}$. 
Our goal is to rewrite the DOS in a way that highlights the fast and slow oscillations, which are not directly evident in Eq.~(\ref{eq:DOSdef}). This is achieved using by the Poisson summation formula \cite{brack97:book,tarasenko02:1769,winkler03:1,ihn10:book} which results in
\begin{eqnarray}
\delta D(B)&\equiv& \frac{D(E_F,B)-D_0}{D_0}  \\
&\simeq&2 \sum_{l=1}^\infty \tilde{L}\left (l \frac{\Gamma}{\hbar \omega_c}\right )\cos(l 2 \pi \mathcal{F}_+ )
\cos(l 2 \pi \mathcal{F}_- ),
\label{eq:dD_F}
\end{eqnarray}
where $D_0=\frac{\hbar^2}{\pi m^*}$ is the zero-field DOS, $\tilde{L}_\Gamma$ is the cosine transform of the broadening function, and the functions $\mathcal{F}_\pm=\frac{1}{2} \left ( F_+ \pm F_- \right )$ represent the fast ($+$) and slow ($-$) parts of the  SdH oscillations, respectively. Details of this derivation are found in App.\ \ref{app:poisson}.
The functions $F_s=F_s(E_F,B)$, with $s=\pm 1$, are defined by the relation
\begin{equation}
    \varepsilon_{n,s}(B)= \frac{E_F}{\hbar \omega_c}
  \Leftrightarrow n=F_s(E_F,B), 
\end{equation}
so determining $F_s$ becomes a root finding problem.  In Fig.\ \ref{fig:Fig4}a) we plot a zoom-in of  $\varepsilon_{n,s}(B)$ along with $E_F/\hbar \omega_c$ [gray solid line]. Accepting {\em non-integer} values of $n$ allows the energy levels to cross $E_F/\hbar \omega_c$ for fixed values of $1/B$ and $E_F$.  The dominant behavior of  $\varepsilon_{n,s}(B)$ with respect to $n$ is linear (see App.\ \ref{app:Rashba}) as is visible in Fig.\ \ref{fig:Fig4}b).  The energy levels cross $E_F/\hbar \omega_c$ at values $n_+$ and $n_-$, for $\varepsilon_{n,+}(B)$ and $\varepsilon_{n,-}(B)$, respectively, which are the values of the corresponding $F_s$-functions : $n_s=F_s(E_F,B)$.

\begin{figure}
\centering
\includegraphics[angle=-0,width=0.48\textwidth]{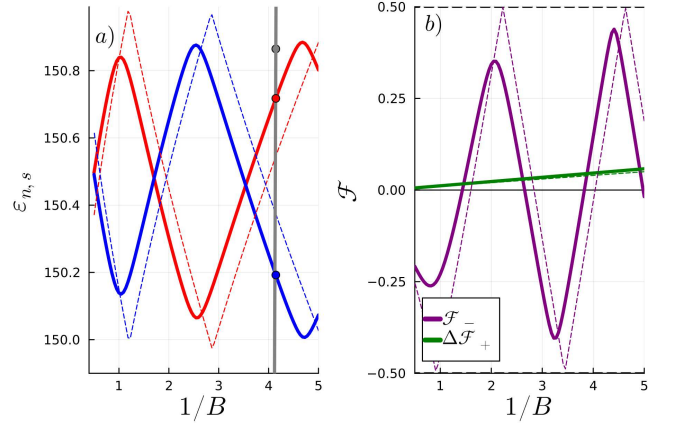}
\caption{a) Energy level $n=150$ using the PD algorithm for $\alpha=7.5$\,meV\,nm and $\beta=3.0$\,meV\,nm.  The dashed curves correspond to $\alpha=7.5$\,meV\,nm and $\beta=0$.  The solid gray line shows $E_F/\hbar \omega_c$, and its value at $1/B=4.145$ [gray circle]. b) The $\mathcal{F}$-function for $\alpha=7.5$\,meV\,nm and $\beta=3.0$\,meV\,nm [solid lines], and pure Rashba, $\beta=0$ [dashed line]. Other parameters are $m^*=0.04$, $g^*=-12$ and $n_\mathrm{2D}=0.0176$\,nm$^{-2}$ for InAs based systems.\cite{beukmann17:241401}}
\label{fig:Fig5}
\end{figure}

As seen in Fig.\ \ref{fig:Fig1}b), gaps open in the spectrum when both $\alpha$ and $\beta$ are non-zero.  In Fig.\ \ref{fig:Fig5}a) a zoom-in of $\varepsilon_{n,s}(B)$ is shown along with $E_F/\hbar \omega_c$ [gray solid line] for $\alpha=7.5$\,meV\,nm and $\beta=3.0$\,meV\,nm. The dashed curves are the corresponding pure Rashba eigenenergies.  
Note the sawtooth shape of the dashed curves since all states cross in this case.  The corresponding $\mathcal{F}$-functions are shown in Fig.\ \ref{fig:Fig5}b).  The anticrossings in the spectrum are visible as a rounding of the sawtooth shape, and level crossings correspond to  $\mathcal{F}_-=0$.

 It is instructive to look at the $\mathcal{F}_\pm$-function in the case of pure Rashba SOI,
\begin{eqnarray}
\mathcal{F}_+&=&\frac{E_F}{\hbar \omega_c}+\frac{2 \varepsilon_R}{\hbar \omega_c}-\frac{1}{2} = \frac{h}{2e} n_{2D} \frac{1}{B}-\frac{1}{2} +\frac{2 \varepsilon_R}{\hbar \omega_c}, \label{eq:FpR}\\
{\cal F}_{-}  &=&
-\frac{1}{2}+\sqrt{\frac{(1 -\tilde{\Delta})^{2}}{4}+\frac{\varepsilon_R E_F}{(\hbar\omega_c)^2} } \approx 
\frac{m^*\alpha k_F}{e \hbar} \frac{1}{B}
\label{eq:FmR},
\end{eqnarray}
where $k_F=\sqrt{2\pi n_\mathrm{2D}}$, and the approximate sign in Eq.\ (\ref{eq:FmR}) refers to the {\em low} field limit.  Since the SdH oscillation frequency in Eq.\ (\ref{eq:FpR}) is dominated by the term proportional to $n_\mathrm{2D}$, we define the spin-orbit related contribution to the fast oscillations as
\begin{equation}
    \Delta \mathcal{F}_+ \equiv  \mathcal{F}_+ - \left ( 
     \frac{h}{2e} n_{2D} \frac{1}{B}-\frac{1}{2}
    \right ) .
\end{equation}
This allows us to plot on the same graph the slow spin-orbit related oscillations described by $\mathcal{F}_-$ and the spin-orbit related modification of the fast oscillation $\Delta \mathcal{F}_+$.
Note that the sawtooth shape in Fig. \ref{fig:Fig5}b) for the case of pure Rashba SOI [purple dashed curve] have a fixed slope $\pm \frac{m^*\alpha k_F}{e \hbar}$.  This is equivalent to the result in Eq.\ (\ref{eq:FmR}), which is linear in $1/B$, since $\cos(2 l \pi \mathcal{F}_-)=\cos(-2 l \pi \mathcal{F}_-)$, i.e.\ the sign of the $\mathcal{F}_-$-slope is irrelevant.

\begin{figure}[hbt]
\centering
\includegraphics[angle=-0,width=0.49\textwidth]{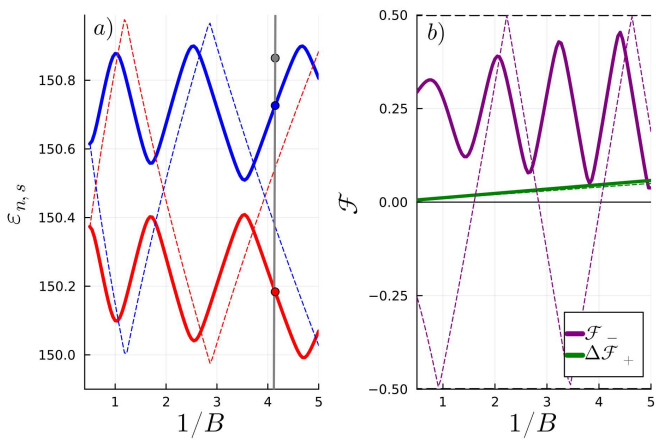}
\caption{a) Energy level $n=150$ using the PD algorithm for $\alpha=7.5$\,meV\,nm, $\beta=3.0$\,meV\,nm, and $\theta=\pi/3$.  The dashed curves correspond to $\alpha=7.5$\,meV\,nm and $\beta=0$.  The solid gray line shows $E_F/\hbar \omega_c$, and its value at $1/B=4.145$ [gray circle]. b) The $\mathcal{F}$-function for $\alpha=7.5$\,meV\,nm and $\beta=3.0$\,meV\,nm [solid lines], and pure Rashba, $\beta=0$ [dashed line]. Other parameters are $m^*=0.04$, $g^*=-12$ and $n_\mathrm{2D}=0.0176$\,nm$^{-2}$ for InAs based systems.\cite{beukmann17:241401}}
\label{fig:Fig6}
\end{figure}
Finally, we consider the influence of an in-plane component of the magnetic field, i.e.\ $\theta \neq 0$.  In this case the parity symmetry no longer holds and {\em all} states anticross as seen in Fig.\ \ref{fig:Fig6}a).  This results in no states simultaneously crossing $E_F$, due to the level repulsion. Note that $\mathcal{F}_-=0$ corresponds to both  pseudo-spin species simultaneously crossing $E_F$ at a given $B$-field.
These new anticrossings have a direct effect on the $\mathcal{F}_-$-function here, which never reaches zero, as opposed to Fig.\ \ref{fig:Fig5}b) where $\mathcal{F}_-$ takes both positive and negative values.
The $\mathcal{F}_-$ thus contains information on how close to (or far from) each other states with opposite $s$ cross $E_F$.  This property is useful when interpreting so-called {\em coincidence measurements} \cite{fang68:823} that have been used to map out level crossings in SdH oscillations in 2DEGs in tilted magnetic fields.\cite{brosig00:13045,hatke12:241305}

\section{Fitting magnetotransport data}
\label{sec:fit}
The oscillation frequencies introduced in the previous section allows for a convenient separation of tasks when  analyzing the magneto-oscillations. In 2D systems, the longitudinal resistance is proportional to the DOS \cite{ihn10:book}, so the previous analysis applies directly to their magneto-oscillations.  The rapid oscillations, i.e.\ SdH oscillation frequency $f^\mathrm{SdH}=\frac{h}{2e}n_\mathrm{2D}$  can be easily extracted by calculating the frequency spectrum via FFT, thus yielding the 2DEG density $n_\mathrm{2D}$ \cite{engels1997experimental,beukmann17:241401}.  The remaining parameters ($\alpha$, $\beta$ and $\Gamma$) can be found by fitting the {\em slow} spin-orbit related oscillations.  We outline below this procedure for fitting realistic magnetoresistance data.  

\begin{figure}
\centering
\includegraphics[angle=-0,width=0.47\textwidth]{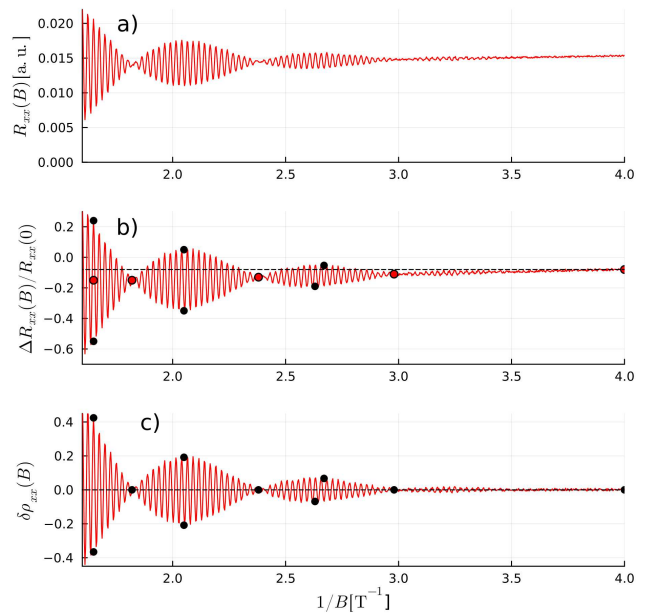}
\caption{a) Magnetoresistance data generated using Eq.\ (\ref{eq:DOSdef}) with $\Gamma=0.45$\,meV, $n_\mathrm{2D}=0.019$\,nm$^{-2}$, $\alpha=7.20$\,meV\,nm and $\beta=2.40$\,meV\,nm. Slight background slope and random noise is to mimic realistic measurements. b) Normalized magneto-oscillations showing slope due to background. Extrema and zeros are indicated by black and red dots, respectively. c) Proper normalized magneto-oscillations after subtracting background [see Sec.\ \ref{sec:fit} for details]. Other parameters are $m^*=0.04$, $g^*=-12$. \cite{beukmann17:241401}}
\label{fig:Fig7}
\end{figure}

Our starting point is Eq.~(\ref{eq:DOSdef}), which we use to generate realistic magnetoresistance data~\cite{beukmann17:241401}.  We use parameters $\Gamma=0.45$\,meV, $n_\mathrm{2D}=0.019$\,nm$^{-2}$, $\alpha=7.20$\,meV\,nm and $\beta=2.40$\,meV\,nm, and add a slight background and noise components to better mimic realistic data. The resulting
$R_{xx}(B)$ is shown in Fig.\ \ref{fig:Fig7}a) where a slight upward slope is barely discernible.  From the $R_{xx}$ data, the normalized magneto-oscillation is calculated
\begin{equation}
    \Delta R_{xx}=\frac{R_{xx}(B)-R_{xx;0}}{R_{xx;0}},
\end{equation}
where $R_{xx;0}$ is defined as the resistance at the magnetic field where the oscillations have been fully suppressed, in this case for $B \leq 0.25$\,T.  This is plotted in Fig.\ \ref{fig:Fig7}b), where the extremas have been marked with black dots, and central points (zeros) are marked with red dots.  The background signal showing a slight upward trend is now more visible. The data is brought to the proper normalized magneto-oscillation form, shown in Fig.\ \ref{fig:Fig7}c), by subtracting the background using a simple linear interpolation between the middle points [red points in Fig.\ \ref{fig:Fig7}b)]. At this point, the data can be directly fitted to the {\em slow} oscillating terms in Eq.\ (\ref{eq:dD_F}) using only a small number of points [black dots].  Due to background compensation we introduce an extra parameter $R_0$, so the resulting {\em slow} envelope function used for fitting is
\begin{equation}
    \delta \rho_{xx}(B)=2 R_0 \tilde{L}_\Gamma \left (
    \frac{\Gamma}{\hbar \omega_c}\right  ) \cos(2 \pi \mathcal{F}_-(B;\alpha,\beta)).
    \label{eq:dD_F_R0}
\end{equation}
\begin{figure}[htb!]
\centering
\includegraphics[angle=-0,width=0.47\textwidth]{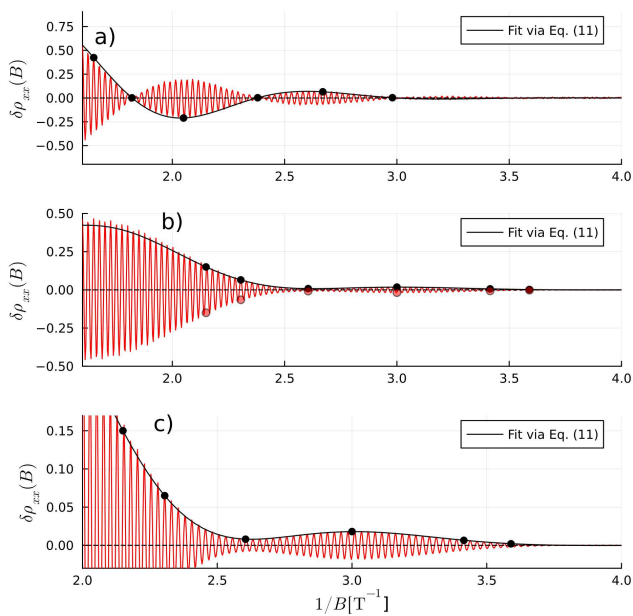}
\caption{a) The normalized magneto-oscillations along with fitted curve through 6 data points [black dots].  Resulting fitted parameters are $\alpha=(7.24 \pm 0.06)$\,meV\,nm and $\beta=(2.5 \pm 0.3)$\,meV\,nm. b) Normalized magneto-oscillations generated using Eq.\ (\ref{eq:DOSdef}) with $\Gamma=0.45$\,meV, $\alpha=3.30$\,meV\,nm and $\beta=5.60$\,meV\,nm [see main text]. c) Zoom in on reference points and fitted curve which yields fit values $\alpha=(3.33 \pm 0.03)$\,meV\,nm and $\beta=(5.63 \pm 0.02)$\,meV\,nm.}
\label{fig:Fig8}
\end{figure}
Fitting the data in Fig.\ \ref{fig:Fig7}c) to Eq.\ (\ref{eq:dD_F_R0}) results in a {\em slow} envelope shown in Fig.\ \ref{fig:Fig8}a).  The results of the fitting yields parameter values $R_0=1.2 \pm 0.2$, $B_q=0.71 \pm 0.03$\,T, $\alpha=(7.24 \pm 0.06)$\,meV\,nm and $\beta=(2.5 \pm 0.3)$\,meV\,nm.  The fitting only takes a few tens of seconds, and a few attempts for finding a good starting point for the fitting parameters. Note that the time to generate the full data took a couple of hours (on the same computer).  
Attempting to fit real transport data using Eq.\ (\ref{eq:DOSdef}), which requires calculating the whole spectrum $\varepsilon_{n,s}(B)$ for all $B$-values to capture both fast {\em and} slow oscillations, would thus be prohibitively time consuming.  Our method circumvents this problem by extracting the important slow spin-orbit-related oscillations via $\mathcal{F}_-(B)$, which are easily fitted using only 5-10 magnetic field points.

Finally, we point out that for cases where Rashba and Dresselhaus SOI parameters are close to each other in value, the slow part of the magnetoscillations does not cross zero, i.e.\ there are no beating nodes \cite{candido23:cond-mat}. This can be seen in magneto-oscillation data in Fig.\ \ref{fig:Fig8}b)  generated using $\alpha=3.30$\,meV\,nm and $\beta=5.60$\,meV\,nm.  The background can be subtracted using center points between the red and black dots.  In Fig.\ \ref{fig:Fig8}c) a zoom-in of the reference points and fitted curve is shown.  The fit values are $\alpha=(3.33 \pm 0.03)$\,meV\,nm and $\beta=(5.63 \pm 0.02)$\,meV\,nm, which is a very good agreement
with the parameter values used to generate the original data.
Note that in both cases of Figs.\ \ref{fig:Fig8}a) and \ref{fig:Fig8}b), the reference points fulfill $\delta \rho_{xx} < 0.4$, which ensures that the higher harmonics can be neglected, due to the exponential suppression \cite{candido23:cond-mat}.

\section{Conclusion} 
\label{sec:Conclusion}
In this paper we presented a new method to efficiently calculate the relevant energy spectrum for SdH magneto-oscillation analysis.  We showed that the numerical procedure along with the Poisson summation formula allow for an efficient calculation and a better understanding of the fast and slow magneto-oscillations.  The spin-orbit parameters $\alpha$ and $\beta$, and the Landau level broadening $\Gamma$, can be uniquely extracted from $\mathcal{F}_-$, which oscillates slowly.  To illustrate our method we applied it to realistic magneto-transport data and find that fitting the slow oscillations yields very quick and accurate fit results. The slow oscillations in $\mathcal{F}_-$ can also shed light on so-called coincidence measurements on tilted magnetic fields. Our method does not rely on finding beating nodes so it can be used to fit data in case of $\alpha$ and $\beta$ being comparable in size.

\section{Acknowledgment}
The authors acknowledge funding from the Reykjavik University PhD Fund, the S\~ao Paulo Research Foundation (FAPESP) Grants No. 2016/08468-0 and No. 2020/00841-9, Conselho Nacional de Pesquisas (CNPq), Grants No. 306122/2018-9 and 301595/2022-4. 

\appendix
\section{$\varepsilon_{n,s}$ and $F_s$ in the case of pure Rashba}
\label{app:Rashba}
The Hamiltonian in Eq.\ (\ref{eq:HfullLadder}) with $\beta=\theta=0$ results in the pure Rashba Hamiltonian
\begin{equation}
    H_R=\left  (a^\dagger a +\frac{1}{2} \right )+\frac{\tilde{\Delta}}{2}\sigma_z +\frac{\alpha}{\sqrt{2}\hbar \omega_c \ell_c} (a^\dagger \sigma_-+ a \sigma_-).
\end{equation}
This can be written in $2\times 2$ subspaces $\{ |n,\uparrow \rangle,|n+1,\downarrow \rangle \}$, $n=0,1, \dots$,\cite{candido23:cond-mat} which results in the matrix 
\begin{eqnarray}
    H_{R;2 \times 2}&=&
    \left [
\begin{array}{cc}
     n+\frac{(1+\tilde{\Delta})}{2}& \frac{2\alpha}{\sqrt{2}\hbar \omega_c \ell_c} (n+1) \\
    \frac{2\alpha}{\sqrt{2}\hbar \omega_c \ell_c} (n+1) & n+1+\frac{(1-\tilde{\Delta})}{2}
\end{array}
    \right ] \nonumber \\
    &=& (n+1)+ \left [
\begin{array}{cc}
     \frac{(1-\tilde{\Delta})}{2}& \frac{\sqrt{2} \alpha}{\hbar \omega_c \ell_c} (n+1) \\
    \frac{\sqrt{2}\alpha}{\hbar \omega_c \ell_c} (n+1) & -\frac{(1-\tilde{\Delta})}{2}
\end{array}
    \right ],
\end{eqnarray}
with eigenvalues
\begin{eqnarray}
\varepsilon_{n,+}&=&(n+1)-\sqrt{\frac{(1-\tilde{\Delta})^2}{4}+\frac{2 m^* \alpha^2}{\hbar \omega_c \hbar^2}(n+1)} \\
\varepsilon_{n+1,-}&=&(n+1)+\sqrt{\frac{(1-\tilde{\Delta})^2}{4}+\frac{2 m^* \alpha^2}{\hbar \omega_c \hbar^2}(n+1)}.
\end{eqnarray}
The above equations reduce to Eqs.\ (\ref{eq:Enp}) and (\ref{eq:Enm}) using $\varepsilon_R=\frac{m^*\alpha}{2\hbar^2}$.  The labelling of the eigenstates is chosen such that in the limit $\alpha \rightarrow 0$ the eigenstates evolve into the correct eigenstates in the {\em absence} of SOI: $\varepsilon_{n,+} \rightarrow \varepsilon^0_{n,\uparrow}$ and $\varepsilon_{n+1,-} \rightarrow \varepsilon^0_{n+1,\downarrow}$.

The definition of the DOS in Eq.\  (\ref{eq:DOSdef}) contains a sum over $n=0,1,2,\dots$ which can be formally written as an integral over the {\em continuous} variable $x$ via the Poisson summation formula (also known as trace formula \cite{brack97:book}) in Eq.\ (\ref{eq:poisson}).  Since the eigenenergies $\varepsilon_{n,s}$ are a well defined function of $n$, the index $n$ can be replaced by a continuous variable $x \in [0, \infty)$. The derivative of the eigenenergies with respect to $x$ can then be calculated
\begin{eqnarray}
    \frac{\partial \varepsilon_{x,+}}{\partial x}
    &=& 1-\frac{\frac{2\varepsilon_R}{\hbar \omega_c}}{\sqrt{\frac{(1-\tilde{\Delta})^2}{4}+\frac{4 \varepsilon_R}{\hbar \omega_c }(x+1)}}, \\
    & \approx&  1-\sqrt{\frac{\varepsilon_R}{E_F}} \approx 1, 
    \label{eq:dedn}
\end{eqnarray}
where we used $x+1 \approx \frac{E_F}{\hbar  \omega_c}$.  The same argument applies to  $\varepsilon_{x,-}$, i.e.\ $\frac{\partial \varepsilon_{x,-}}{\partial x}\approx 1$.
In the case of non-zero $\beta$ and/or $\theta$ in Eq.\ (\ref{eq:HfullLadder})
will lead to anticrossings, which tend to {\em flatten} the square root behavior of the energy levels, see Fig.\ \ref{fig:Fig1}b) and c), thus making the approximation in Eq.\ (\ref{eq:dedn}) even better. 

\section{Partial Hamiltonians and parity}
Here we describe the form of the partial Hamiltonian in the case of parity symmetry \cite{candido23:cond-mat}, and in the absence of that symmetry.

\subsection{Matrix elements and parity symmetry}
\label{app:HPD}
As was outlined in Ref.\ \onlinecite{candido23:cond-mat} the full basis $|m,\sigma \rangle$ can be split in two according to the eigenvalues of the parity operator
\begin{equation}
\hat{\mathcal{P}}=\exp(i\pi (a^{\dagger}a+\sfrac{1}{2}(\sigma_{z}-1))),
\end{equation}
which are $\mathcal{P}= \pm 1$.  The basis states of the resulting parity subspace  are then
\begin{eqnarray}
\mathcal{P} =+1&:& \quad \{ |0, \uparrow \rangle, |1, \downarrow \rangle,  |2, \uparrow \rangle, |3, \downarrow \rangle, |4, \uparrow \rangle, \dots \}  \label{eq:Pp} \\
\mathcal{P} =-1&:& \quad \{ |0, \downarrow \rangle, |1, \uparrow \rangle,  |2, \downarrow \rangle, |3, \uparrow \rangle, |4, \downarrow \rangle, \dots \} \label{eq:Pm} .
\end{eqnarray}
The Hamiltonian matrix for each $\mathcal{P}= \pm 1$ subspace become a tridiagonal matrix with diagonal elements
\begin{eqnarray}
\bigl [ H_\mathrm{2D}^{(+1)} \bigr ]_{k,k}&=&k+\frac{1}{2}+\frac{\tilde{\Delta}}{2}(-1)^k \label{eq:HPpkk}  \\
\bigl [ H_\mathrm{2D}^{(-1)} \bigr ] _{k,k}&=&k+\frac{1}{2}-\frac{\tilde{\Delta}}{2}(-1)^k, \label{eq:HPmkk} 
\end{eqnarray}
where $k=0,1,2,\dots$ labels the basis states in subspace $\mathcal{P}= \pm 1$.  The alternating sign of the Zeeman term reflects the alternating $\uparrow$ and $\downarrow$ in the basis states in Eq.\ (\ref{eq:Pp}) and (\ref{eq:Pm}).  The off-diagonal matrix elements are given by
\begin{eqnarray}
\Bigl [ H_\mathrm{2D}^{(+1)} \Bigr ]_{k,k+1} &=& \sqrt{k+1} \Bigl ( \frac{i \alpha}{\sqrt{2}\hbar \omega_c \ell_c}  (1+(-1)^k) \nonumber \\
& &+\frac{\beta}{\sqrt{2}\hbar \omega_c \ell_c} (1-(-1)^k) \Bigr ) ,   \label{eq:HPpkk1} \\
\Bigl [ H_{2D}^{(-1)} \Bigr ]_{k,k+1} &=&\sqrt{k+1} \Bigl ( \frac{i \alpha}{\sqrt{2}\hbar \omega_c \ell_c}  (1-(-1)^k) \nonumber \\
& &+\frac{\beta}{\sqrt{2}\hbar \omega_c \ell_c} (1+(-1)^k) \Bigr ) .
    \label{eq:HPmkk1}
\end{eqnarray}
The $(1 \pm (-1)^{k})$ terms take alternating values 0 and 2, 
which results in $\Bigl [ H_\mathrm{2D}^{(+1)} \Bigr ]_{k,k+1} \propto \alpha$ for even values of $k$, but for odd value of $k$ gives $\Bigl [ H_\mathrm{2D}^{(+1)} \Bigr ]_{k,k+1} \propto \beta$.  For the $\mathcal{P}=-1$ parity subspace, the matrix elements  in Eq.\ (\ref{eq:HPmkk1}), the even/odd pattern for $k$ is switched. 
Compare this to Eq.\ (\ref{eq:H2Dmm1}) in the {\em absence} of parity symmetry where each $2 \times 2$ block contains both $\alpha$ and $\beta$.

The partial Hamiltonian for $\mathcal{P}=+1$ centered on the $n$-th Landau level is constructed from Eqs.\ (\ref{eq:HPpkk}) and (\ref{eq:HPpkk1})
\begin{eqnarray}
    \Bigl [H_\mathrm{PD}^{(+1)}(n) \Bigr]_{m,m}&=& \left (n+(m-N_\mathrm{PD}-1)+\frac{1}{2} \right ) \nonumber \\
    & &+\frac{\tilde{\Delta}}{2}(-1)^{(\bar{n}+m-N_\mathrm{PD}-1)}, \\
\Bigl [ H_\mathrm{PD}^{(+1)} (n)\Bigr ]_{m,m+1} \nonumber 
    &=& \sqrt{(n+m-N_\mathrm{PD}-1)+1}  \times \nonumber \\
    & &  \Bigl (  \frac{i \alpha}{\sqrt{2}\hbar \omega_c \ell_c} (1+(-1)^{(\bar{n} +m-N_\mathrm{PD}-1)})   \nonumber \\
    &+& \frac{\beta}{\sqrt{2}\hbar \omega_c \ell_c}   (1-(-1)^{(\bar{n} +m-N_\mathrm{PD}-1)}) \Bigr ) \nonumber \\, 
\end{eqnarray}
where $\bar{n}=\mathrm{round}(n)$ and $m \in [1,2N_\mathrm{PD}+1]$.  The eigen\-energy $\varepsilon^{(+1)}_{n,s}$ is obtained as the $(N_\mathrm{PD}+1)$-th eigenvalue of $H_\mathrm{PD}^{(+1)}(n)$.
In a similar fashion the partial Hamiltonian for $\mathcal{P}=-1$, centered on $n$-th Landau level, is given by
\begin{eqnarray}
    \Bigl [H_\mathrm{PD}^{(-1)}(n) \Bigr]_{m,m}&=& \left (n+(m-N_\mathrm{PD}-1)+\frac{1}{2} \right ) \nonumber \\
    & &+\frac{\tilde{\Delta}}{2}(-1)^{(\bar{n}+m-N_\mathrm{PD}-1)}, \\
\Bigl [ H_\mathrm{PD}^{(-1)} (n)\Bigr ]_{m,m+1} \nonumber 
    &=& \sqrt{(n+m-N_\mathrm{PD}-1)+1}  \times \nonumber \\
    & &  \Bigl (  \frac{i \alpha}{\sqrt{2}\hbar \omega_c \ell_c} (1-(-1)^{(\bar{n} +m-N_\mathrm{PD}-1)})   \nonumber \\
    &+& \frac{\beta}{\sqrt{2}\hbar \omega_c \ell_c}   (1+(-1)^{(\bar{n} +m-N_\mathrm{PD}-1)}) \Bigr ), \nonumber \\ 
\end{eqnarray}
and the eigen\-energy $\varepsilon^{(-1)}_{n,s}$, is obtained as the $(N_\mathrm{PD}+1)$-th eigenvalue of $H_\mathrm{PD}^{(-1)}(n)$.

\subsection{Matrix elements {\em without} parity symmetry}
\label{app:noParity}
In the absence of parity symmetry, i.e.\ for $\theta \neq 0$, the matrix elements in Eqs.\ (\ref{eq:H2Dmm}) and (\ref{eq:H2Dmm1}) are used to construct the partial matrix centered on the $n$-th Landau level
\begin{eqnarray}
    \Bigl [H_\mathrm{PD}(n) \Bigr]_{m,m}&=&
    \Bigl ( (n+m-N_\mathrm{PD}-1)  +\frac{1}{2}\Bigr ) \left [
    \begin{array}{cc}
     1 & 0  \\
     0  & 1 
    \end{array}
    \right ]  \nonumber \\
    & +& 
    \left [
    \begin{array}{cc}
     \frac{\tilde{\Delta}}{2} & \frac{\tilde{\Delta}}{2}\tan(\theta)e^{i \phi}  \\
     \frac{\tilde{\Delta}}{2}\tan(\theta)e^{-i \phi}  &  - \frac{\tilde{\Delta}}{2} 
    \end{array}
    \right ], \\
\Bigl [ H_\mathrm{PD} (n)\Bigr ]_{m,m+1} 
    &=& \sqrt{(n+m-N_\mathrm{PD}-1)+1}  \times  \nonumber \\
    & &\frac{1}{\sqrt{2}\hbar \omega_c \ell_c}
    \left [
    \begin{array}{cc}
     0 & 2\beta \\
     -2i \alpha & 0
    \end{array}
    \right ],
\end{eqnarray}
where $m \in [1,2N_\mathrm{PD}+1]$.
The eigenenergy pair $\varepsilon_{n,+}$ and $\varepsilon_{n,-}$ are obtained as eigenvalues of $H_\mathrm{PD}(n)$ number  $(2N_\mathrm{PD}+1)$ and $(2N_\mathrm{PD}+2)$. 

\section{Poisson's summation formula}
\label{app:poisson}
Here we will apply the Poisson summation formula
\begin{equation}
    \sum_{n=0}^\infty f(n)= \int_0^\infty dx f(x)+2 \sum_{l=1}^\infty \int_0^\infty dx f(x) \cos(l 2 \pi x),
    \label{eq:poisson}
\end{equation}
to the sum over the broadened Landau levels in Eq.\~(\ref{eq:DOSdef}).  Starting with one spin species $s$
\begin{eqnarray}
&&\sum_{n=0}^\infty L_\Gamma(E_F-\hbar \omega_c \varepsilon_{n,s}(B))    \\
&=& \int_0^\infty dx L_\Gamma(E_F-\hbar \omega_c \varepsilon_{x,s}(B))  \nonumber \\
&+& \sum_{l=1}^\infty \int_0^\infty dx L_\Gamma(E_F-\hbar \omega_c \varepsilon_{x,s}(B)) \cos(l 2\pi x).
\end{eqnarray}
Next, we introduce a change of variables
\begin{eqnarray}
u&=&E_F-\hbar \omega_c \varepsilon_{x,s}(B) , \\
\frac{du}{dx}&=&-\hbar \omega_c \frac{\partial \varepsilon_{x,s}}{\partial x}.
\end{eqnarray}
The derivative $\frac{\partial \varepsilon_{x,s}}{\partial x} =1+\mathcal{O}(\sqrt{\varepsilon_R/E_F})$ when evaluated at $\varepsilon_{x,s} \approx E_F/\hbar \omega_c$
\begin{eqnarray}
    &&\sum_{n=0}^\infty L_\Gamma(E_F-\hbar \omega_c \varepsilon_{n,s}(B))    \\
&=&  \frac{1}{\hbar \omega_c} \left (\int_{-\infty}^\infty du L_\Gamma(u)  \right .\nonumber \\
&+&  \left . 2\sum_{l=1}^\infty \int_{-\infty}^\infty du L_\Gamma(u) \cos \left (l 2\pi F_s(E_F-u,B) \right ) \right ).
\label{eq:Fint}
\end{eqnarray}
In order to keep the equations as concise as possible, we will now drop the $B$ argument in both $F_s$ and $\varepsilon_{x,s}$.
The integrand in Eq.\ (\ref{eq:Fint}) has width $\sim \Gamma$, and since $E_F \gg \Gamma$, we can use 1st order Taylor expansion of the $F_s$ function in terms of $u$
\begin{eqnarray}
 F_s(E_F-u) & = & F_s(E_F)-F'(E_F) u + \mathcal{O}(u^2)  \\
 &\approx & F_s(E_F)-\frac{1}{\hbar \omega_c} u ,
 \label{eq:FsTaylor}
\end{eqnarray}
where we have used $\frac{ d F_s(E_F)}{d E_F}=\frac{1}{\hbar \omega_c}$, which is a consequence of $\frac{\partial \varepsilon_{n,s}}{\partial n} =1$. This can be shown using that $n=F_s(E_F)$ is the inverse function of $E_F=\hbar \omega_c \varepsilon_{n,s}$, i.e.\ $n=F_s(\hbar \omega_c \varepsilon_{n,s})$.  Taking the derivative of this relation with respect to $n$ results in
\begin{equation}
1=\frac{dF_s(E_F)}{dE_F}\hbar \omega_c \frac{\partial \varepsilon_{n,s}}{\partial n} \approx  \frac{dF_s(E_F)}{dE_F}\hbar \omega_c,
\end{equation}
which yields the relation below Eq.\ (\ref{eq:FsTaylor}).
We can thus write Eq.~(\ref{eq:Fint}) as
\begin{eqnarray}
    &&\sum_{n=0}^\infty L_\Gamma(E_F-\hbar \omega_c \varepsilon_{n,s})    \\
&=&  \frac{1}{\hbar \omega_c} \left (\int_{-\infty}^\infty du L_\Gamma(u)  \right .\nonumber \\
&+& \left . 2\sum_{l=1}^\infty  \cos(l 2 \pi F_s(E_F)) \int_{-\infty}^\infty du L_\Gamma(u) \cos \left (l 2\pi \frac{u}{\hbar  \omega_c} \right ) \right ) \nonumber \\
&=& \frac{1}{\hbar \omega_c} \left (1  + 2\sum_{l=1}^\infty  \cos(l 2 \pi F_s(E_F,B))  \tilde{L}_\Gamma \left (l \frac{\Gamma}{\hbar \omega_c} \right ) \right ), \nonumber \\
\end{eqnarray}
where the symmetric broadening will make the sine-term appearing in the Taylor expansion vanish.  The cosine transform is defined as
\begin{equation}
    \tilde{L}_\Gamma\left (l \frac{\Gamma}{\hbar \omega_c} \right )=\int_{-\infty}^\infty du L_\Gamma(u) \cos \left (l 2\pi \frac{u}{\hbar \omega_c} \right ),
\end{equation}
which, for Gaussian broadening, leads to
\begin{equation}
    \tilde{L}_\Gamma\left (l \frac{\Gamma}{\hbar \omega_c} \right )=\exp{ \left (
-\left [ \sqrt{2}\pi \, l\frac{ \Gamma}{\hbar \omega_c} \right ]^2
    \right )}=e^{-l^2\frac{B_q^2}{B^2}},
\end{equation}
where we $B_q=\sqrt{2}\pi \frac{m^*\Gamma}{\hbar e}$.
Finally, applying this to Eq.~(\ref{eq:DOSdef}) and using the trigonometric relation
\begin{eqnarray}
    & &\cos( l 2\pi F_+)+\cos( l 2\pi F_+) \\
    &=&2 \cos \left (l 2 \pi  \frac{F_+ + F_-}{2} \right) \cos \left (l 2 \pi  \frac{F_+ - F_-}{2} \right),
\end{eqnarray}
and $\frac{1}{2 \pi \ell_c^2}\frac{1}{\hbar \omega_c}=\frac{\hbar^2}{2 \pi m^*}$
results in Eq.\ (\ref{eq:dD_F}).
 

\begin{thebibliography}{31}%
\makeatletter
\providecommand \@ifxundefined [1]{%
 \@ifx{#1\undefined}
}%
\providecommand \@ifnum [1]{%
 \ifnum #1\expandafter \@firstoftwo
 \else \expandafter \@secondoftwo
 \fi
}%
\providecommand \@ifx [1]{%
 \ifx #1\expandafter \@firstoftwo
 \else \expandafter \@secondoftwo
 \fi
}%
\providecommand \natexlab [1]{#1}%
\providecommand \enquote  [1]{``#1''}%
\providecommand \bibnamefont  [1]{#1}%
\providecommand \bibfnamefont [1]{#1}%
\providecommand \citenamefont [1]{#1}%
\providecommand \href@noop [0]{\@secondoftwo}%
\providecommand \href [0]{\begingroup \@sanitize@url \@href}%
\providecommand \@href[1]{\@@startlink{#1}\@@href}%
\providecommand \@@href[1]{\endgroup#1\@@endlink}%
\providecommand \@sanitize@url [0]{\catcode `\\12\catcode `\$12\catcode
  `\&12\catcode `\#12\catcode `\^12\catcode `\_12\catcode `\%12\relax}%
\providecommand \@@startlink[1]{}%
\providecommand \@@endlink[0]{}%
\providecommand \url  [0]{\begingroup\@sanitize@url \@url }%
\providecommand \@url [1]{\endgroup\@href {#1}{\urlprefix }}%
\providecommand \urlprefix  [0]{URL }%
\providecommand \Eprint [0]{\href }%
\providecommand \doibase [0]{http://dx.doi.org/}%
\providecommand \selectlanguage [0]{\@gobble}%
\providecommand \bibinfo  [0]{\@secondoftwo}%
\providecommand \bibfield  [0]{\@secondoftwo}%
\providecommand \translation [1]{[#1]}%
\providecommand \BibitemOpen [0]{}%
\providecommand \bibitemStop [0]{}%
\providecommand \bibitemNoStop [0]{.\EOS\space}%
\providecommand \EOS [0]{\spacefactor3000\relax}%
\providecommand \BibitemShut  [1]{\csname bibitem#1\endcsname}%
\let\auto@bib@innerbib\@empty
\bibitem [{\citenamefont {Shubnikov}\ and\ \citenamefont
  {de~Haas}(1930{\natexlab{a}})}]{sdh-original}%
  \BibitemOpen
  \bibfield  {author} {\bibinfo {author} {\bibfnamefont {L.}~\bibnamefont
  {Shubnikov}}\ and\ \bibinfo {author} {\bibfnamefont {W.}~\bibnamefont
  {de~Haas}},\ }\href@noop {} {\bibfield  {journal} {\bibinfo  {journal} {207a,
  b, c, 210a}\ } (\bibinfo {year} {1930}{\natexlab{a}})}\BibitemShut {NoStop}%
\bibitem [{\citenamefont {Shubnikov}\ and\ \citenamefont
  {de~Haas}(1930{\natexlab{b}})}]{sdh-original2}%
  \BibitemOpen
  \bibfield  {author} {\bibinfo {author} {\bibfnamefont {L.}~\bibnamefont
  {Shubnikov}}\ and\ \bibinfo {author} {\bibfnamefont {W.}~\bibnamefont
  {de~Haas}},\ }in\ \href@noop {} {\emph {\bibinfo {booktitle} {Proc.
  Netherlands Roy. Acad. Sci}}},\ Vol.~\bibinfo {volume} {33}\ (\bibinfo {year}
  {1930})\ p.\ \bibinfo {pages} {363}\BibitemShut {NoStop}%
\bibitem [{\citenamefont {Ihn}(2010)}]{ihn10:book}%
  \BibitemOpen
  \bibfield  {author} {\bibinfo {author} {\bibfnamefont {T.}~\bibnamefont
  {Ihn}},\ }\href@noop {} {\emph {\bibinfo {title} {Semiconductor
  Nanostructures}}}\ (\bibinfo  {publisher} {Oxford University Press},\
  \bibinfo {year} {2010})\BibitemShut {NoStop}%
\bibitem [{\citenamefont {Winkler}(2003)}]{winkler03:1}%
  \BibitemOpen
  \bibfield  {author} {\bibinfo {author} {\bibfnamefont {R.}~\bibnamefont
  {Winkler}},\ }\href@noop {} {\emph {\bibinfo {title} {Spin-Orbit Coupling
  Effects in Two-Dimensional Electron and Hole Systems}}}\ (\bibinfo
  {publisher} {Springer Verlag},\ \bibinfo {year} {2003})\BibitemShut {NoStop}%
\bibitem [{\citenamefont {Das}\ \emph {et~al.}(1989)\citenamefont {Das},
  \citenamefont {Miller}, \citenamefont {Datta}, \citenamefont {Reifenberger},
  \citenamefont {Hong}, \citenamefont {Bhattacharya}, \citenamefont {Singh},\
  and\ \citenamefont {Jaffe}}]{das1989evidence}%
  \BibitemOpen
  \bibfield  {author} {\bibinfo {author} {\bibfnamefont {B.}~\bibnamefont
  {Das}}, \bibinfo {author} {\bibfnamefont {D.}~\bibnamefont {Miller}},
  \bibinfo {author} {\bibfnamefont {S.}~\bibnamefont {Datta}}, \bibinfo
  {author} {\bibfnamefont {R.}~\bibnamefont {Reifenberger}}, \bibinfo {author}
  {\bibfnamefont {W.}~\bibnamefont {Hong}}, \bibinfo {author} {\bibfnamefont
  {P.}~\bibnamefont {Bhattacharya}}, \bibinfo {author} {\bibfnamefont
  {J.}~\bibnamefont {Singh}}, \ and\ \bibinfo {author} {\bibfnamefont
  {M.}~\bibnamefont {Jaffe}},\ }\href@noop {} {\bibfield  {journal} {\bibinfo
  {journal} {Physical Review B}\ }\textbf {\bibinfo {volume} {39}},\ \bibinfo
  {pages} {1411} (\bibinfo {year} {1989})}\BibitemShut {NoStop}%
\bibitem [{\citenamefont {Bychkov}\ and\ \citenamefont
  {Rashba}(1984)}]{bychkov1984oscillatory}%
  \BibitemOpen
  \bibfield  {author} {\bibinfo {author} {\bibfnamefont {Y.~A.}\ \bibnamefont
  {Bychkov}}\ and\ \bibinfo {author} {\bibfnamefont {E.~I.}\ \bibnamefont
  {Rashba}},\ }\href@noop {} {\bibfield  {journal} {\bibinfo  {journal}
  {Journal of physics C: Solid state physics}\ }\textbf {\bibinfo {volume}
  {17}},\ \bibinfo {pages} {6039} (\bibinfo {year} {1984})}\BibitemShut
  {NoStop}%
\bibitem [{\citenamefont {Das}\ \emph {et~al.}(1990{\natexlab{a}})\citenamefont
  {Das}, \citenamefont {Datta},\ and\ \citenamefont {Reifenberg}}]{das90:8278}%
  \BibitemOpen
  \bibfield  {author} {\bibinfo {author} {\bibfnamefont {B.}~\bibnamefont
  {Das}}, \bibinfo {author} {\bibfnamefont {S.}~\bibnamefont {Datta}}, \ and\
  \bibinfo {author} {\bibfnamefont {R.}~\bibnamefont {Reifenberg}},\
  }\href@noop {} {\bibfield  {journal} {\bibinfo  {journal} {Phys.\ Rev.\ B}\
  }\textbf {\bibinfo {volume} {41}},\ \bibinfo {pages} {8278} (\bibinfo {year}
  {1990}{\natexlab{a}})}\BibitemShut {NoStop}%
\bibitem [{\citenamefont {Das}\ \emph {et~al.}(1990{\natexlab{b}})\citenamefont
  {Das}, \citenamefont {Datta},\ and\ \citenamefont
  {Reifenberger}}]{das1990zero}%
  \BibitemOpen
  \bibfield  {author} {\bibinfo {author} {\bibfnamefont {B.}~\bibnamefont
  {Das}}, \bibinfo {author} {\bibfnamefont {S.}~\bibnamefont {Datta}}, \ and\
  \bibinfo {author} {\bibfnamefont {R.}~\bibnamefont {Reifenberger}},\
  }\href@noop {} {\bibfield  {journal} {\bibinfo  {journal} {Physical Review
  B}\ }\textbf {\bibinfo {volume} {41}},\ \bibinfo {pages} {8278} (\bibinfo
  {year} {1990}{\natexlab{b}})}\BibitemShut {NoStop}%
\bibitem [{\citenamefont {Nitta}\ \emph {et~al.}(1997)\citenamefont {Nitta},
  \citenamefont {Akazaki}, \citenamefont {Takayanagi},\ and\ \citenamefont
  {Enoki}}]{nitta97:1335}%
  \BibitemOpen
  \bibfield  {author} {\bibinfo {author} {\bibfnamefont {J.}~\bibnamefont
  {Nitta}}, \bibinfo {author} {\bibfnamefont {T.}~\bibnamefont {Akazaki}},
  \bibinfo {author} {\bibfnamefont {H.}~\bibnamefont {Takayanagi}}, \ and\
  \bibinfo {author} {\bibfnamefont {T.}~\bibnamefont {Enoki}},\ }\href
  {\doibase 10.1103/PhysRevLett.78.1335} {\bibfield  {journal} {\bibinfo
  {journal} {Phys. Rev. Lett.}\ }\textbf {\bibinfo {volume} {78}},\ \bibinfo
  {pages} {1335} (\bibinfo {year} {1997})}\BibitemShut {NoStop}%
\bibitem [{\citenamefont {Engels}\ \emph {et~al.}(1997)\citenamefont {Engels},
  \citenamefont {Lange}, \citenamefont {Sch{\"a}pers},\ and\ \citenamefont
  {L{\"u}th}}]{engels1997experimental}%
  \BibitemOpen
  \bibfield  {author} {\bibinfo {author} {\bibfnamefont {G.}~\bibnamefont
  {Engels}}, \bibinfo {author} {\bibfnamefont {J.}~\bibnamefont {Lange}},
  \bibinfo {author} {\bibfnamefont {T.}~\bibnamefont {Sch{\"a}pers}}, \ and\
  \bibinfo {author} {\bibfnamefont {H.}~\bibnamefont {L{\"u}th}},\ }\href@noop
  {} {\bibfield  {journal} {\bibinfo  {journal} {Physical Review B}\ }\textbf
  {\bibinfo {volume} {55}},\ \bibinfo {pages} {R1958} (\bibinfo {year}
  {1997})}\BibitemShut {NoStop}%
\bibitem [{\citenamefont {Sch\"{a}pers}\ \emph {et~al.}(1998)\citenamefont
  {Sch\"{a}pers}, \citenamefont {Engels}, \citenamefont {Lange}, \citenamefont
  {Klocke}, \citenamefont {Hollfelder},\ and\ \citenamefont
  {L\"{u}th}}]{schaepers98:4324}%
  \BibitemOpen
  \bibfield  {author} {\bibinfo {author} {\bibfnamefont {T.}~\bibnamefont
  {Sch\"{a}pers}}, \bibinfo {author} {\bibfnamefont {G.}~\bibnamefont
  {Engels}}, \bibinfo {author} {\bibfnamefont {J.}~\bibnamefont {Lange}},
  \bibinfo {author} {\bibfnamefont {T.}~\bibnamefont {Klocke}}, \bibinfo
  {author} {\bibfnamefont {M.}~\bibnamefont {Hollfelder}}, \ and\ \bibinfo
  {author} {\bibfnamefont {H.}~\bibnamefont {L\"{u}th}},\ }\href {\doibase
  10.1063/1.367192} {\bibfield  {journal} {\bibinfo  {journal} {Journal of
  Applied Physics}\ }\textbf {\bibinfo {volume} {83}},\ \bibinfo {pages} {4324}
  (\bibinfo {year} {1998})},\ \Eprint
  {http://arxiv.org/abs/https://pubs.aip.org/aip/jap/article-pdf/83/8/4324/10592945/4324\_1\_online.pdf}
  {https://pubs.aip.org/aip/jap/article-pdf/83/8/4324/10592945/4324\_1\_online.pdf}
  \BibitemShut {NoStop}%
\bibitem [{\citenamefont {Dresselhaus}(1955)}]{dresselhaus1955spin}%
  \BibitemOpen
  \bibfield  {author} {\bibinfo {author} {\bibfnamefont {G.}~\bibnamefont
  {Dresselhaus}},\ }\href@noop {} {\bibfield  {journal} {\bibinfo  {journal}
  {Physical Review}\ }\textbf {\bibinfo {volume} {100}},\ \bibinfo {pages}
  {580} (\bibinfo {year} {1955})}\BibitemShut {NoStop}%
\bibitem [{\citenamefont {Gilbertson}\ \emph {et~al.}(2008)\citenamefont
  {Gilbertson}, \citenamefont {Fearn}, \citenamefont {Jefferson}, \citenamefont
  {Murdin}, \citenamefont {Buckle},\ and\ \citenamefont
  {Cohen}}]{gilbertson2008zero}%
  \BibitemOpen
  \bibfield  {author} {\bibinfo {author} {\bibfnamefont {A.}~\bibnamefont
  {Gilbertson}}, \bibinfo {author} {\bibfnamefont {M.}~\bibnamefont {Fearn}},
  \bibinfo {author} {\bibfnamefont {J.}~\bibnamefont {Jefferson}}, \bibinfo
  {author} {\bibfnamefont {B.}~\bibnamefont {Murdin}}, \bibinfo {author}
  {\bibfnamefont {P.~D.}\ \bibnamefont {Buckle}}, \ and\ \bibinfo {author}
  {\bibfnamefont {L.}~\bibnamefont {Cohen}},\ }\href@noop {} {\bibfield
  {journal} {\bibinfo  {journal} {Physical Review B}\ }\textbf {\bibinfo
  {volume} {77}},\ \bibinfo {pages} {165335} (\bibinfo {year}
  {2008})}\BibitemShut {NoStop}%
\bibitem [{\citenamefont {Akabori}\ \emph {et~al.}(2006)\citenamefont
  {Akabori}, \citenamefont {Sunouchi}, \citenamefont {Kakegawa}, \citenamefont
  {Sato}, \citenamefont {Suzuki},\ and\ \citenamefont
  {Yamada}}]{akabori2006spin}%
  \BibitemOpen
  \bibfield  {author} {\bibinfo {author} {\bibfnamefont {M.}~\bibnamefont
  {Akabori}}, \bibinfo {author} {\bibfnamefont {T.}~\bibnamefont {Sunouchi}},
  \bibinfo {author} {\bibfnamefont {T.}~\bibnamefont {Kakegawa}}, \bibinfo
  {author} {\bibfnamefont {T.}~\bibnamefont {Sato}}, \bibinfo {author}
  {\bibfnamefont {T.-k.}\ \bibnamefont {Suzuki}}, \ and\ \bibinfo {author}
  {\bibfnamefont {S.}~\bibnamefont {Yamada}},\ }\href@noop {} {\bibfield
  {journal} {\bibinfo  {journal} {Physica E: Low-dimensional Systems and
  Nanostructures}\ }\textbf {\bibinfo {volume} {34}},\ \bibinfo {pages} {413}
  (\bibinfo {year} {2006})}\BibitemShut {NoStop}%
\bibitem [{\citenamefont {Averkiev}\ \emph {et~al.}(2005)\citenamefont
  {Averkiev}, \citenamefont {Glazov},\ and\ \citenamefont
  {Tarasenko}}]{averkiev05:543}%
  \BibitemOpen
  \bibfield  {author} {\bibinfo {author} {\bibfnamefont {N.}~\bibnamefont
  {Averkiev}}, \bibinfo {author} {\bibfnamefont {M.}~\bibnamefont {Glazov}}, \
  and\ \bibinfo {author} {\bibfnamefont {S.}~\bibnamefont {Tarasenko}},\
  }\href@noop {} {\bibfield  {journal} {\bibinfo  {journal} {Solid State
  Commun.}\ }\textbf {\bibinfo {volume} {133}},\ \bibinfo {pages} {543}
  (\bibinfo {year} {2005})}\BibitemShut {NoStop}%
\bibitem [{\citenamefont {Tarasenko}\ and\ \citenamefont
  {Averkiev}(2002)}]{tarasenko02:552}%
  \BibitemOpen
  \bibfield  {author} {\bibinfo {author} {\bibfnamefont {S.}~\bibnamefont
  {Tarasenko}}\ and\ \bibinfo {author} {\bibfnamefont {N.}~\bibnamefont
  {Averkiev}},\ }\href@noop {} {\bibfield  {journal} {\bibinfo  {journal} {JETP
  Lett.}\ }\textbf {\bibinfo {volume} {75}},\ \bibinfo {pages} {552} (\bibinfo
  {year} {2002})}\BibitemShut {NoStop}%
\bibitem [{\citenamefont {Tarasenko}(2002)}]{tarasenko02:1769}%
  \BibitemOpen
  \bibfield  {author} {\bibinfo {author} {\bibfnamefont {S.}~\bibnamefont
  {Tarasenko}},\ }\href@noop {} {\bibfield  {journal} {\bibinfo  {journal}
  {Physics of Solid State}\ }\textbf {\bibinfo {volume} {44}},\ \bibinfo
  {pages} {1769} (\bibinfo {year} {2002})}\BibitemShut {NoStop}%
\bibitem [{\citenamefont {Yang}\ and\ \citenamefont
  {Chang}(2006)}]{yang06:045303}%
  \BibitemOpen
  \bibfield  {author} {\bibinfo {author} {\bibfnamefont {W.}~\bibnamefont
  {Yang}}\ and\ \bibinfo {author} {\bibfnamefont {K.}~\bibnamefont {Chang}},\
  }\href@noop {} {\bibfield  {journal} {\bibinfo  {journal} {Phys. Rev. B}\
  }\textbf {\bibinfo {volume} {73}},\ \bibinfo {pages} {045303} (\bibinfo
  {year} {2006})}\BibitemShut {NoStop}%
\bibitem [{\citenamefont {Beukman}\ \emph {et~al.}(2017)\citenamefont
  {Beukman}, \citenamefont {de~Vries}, \citenamefont {van Veen}, \citenamefont
  {Skolasinski}, \citenamefont {Wimmer}, \citenamefont {Qu}, \citenamefont
  {de~Vries}, \citenamefont {Nguyen}, \citenamefont {Yi}, \citenamefont
  {Kiselev}, \citenamefont {Sokolich}, \citenamefont {Manfra}, \citenamefont
  {Nichele}, \citenamefont {Marcus},\ and\ \citenamefont
  {Kouwenhoven}}]{beukmann17:241401}%
  \BibitemOpen
  \bibfield  {author} {\bibinfo {author} {\bibfnamefont {A.~J.~A.}\
  \bibnamefont {Beukman}}, \bibinfo {author} {\bibfnamefont {F.~K.}\
  \bibnamefont {de~Vries}}, \bibinfo {author} {\bibfnamefont {J.}~\bibnamefont
  {van Veen}}, \bibinfo {author} {\bibfnamefont {R.}~\bibnamefont
  {Skolasinski}}, \bibinfo {author} {\bibfnamefont {M.}~\bibnamefont {Wimmer}},
  \bibinfo {author} {\bibfnamefont {F.}~\bibnamefont {Qu}}, \bibinfo {author}
  {\bibfnamefont {D.~T.}\ \bibnamefont {de~Vries}}, \bibinfo {author}
  {\bibfnamefont {B.-M.}\ \bibnamefont {Nguyen}}, \bibinfo {author}
  {\bibfnamefont {W.}~\bibnamefont {Yi}}, \bibinfo {author} {\bibfnamefont
  {A.~A.}\ \bibnamefont {Kiselev}}, \bibinfo {author} {\bibfnamefont
  {M.}~\bibnamefont {Sokolich}}, \bibinfo {author} {\bibfnamefont {M.~J.}\
  \bibnamefont {Manfra}}, \bibinfo {author} {\bibfnamefont {F.}~\bibnamefont
  {Nichele}}, \bibinfo {author} {\bibfnamefont {C.~M.}\ \bibnamefont {Marcus}},
  \ and\ \bibinfo {author} {\bibfnamefont {L.~P.}\ \bibnamefont
  {Kouwenhoven}},\ }\href@noop {} {\bibfield  {journal} {\bibinfo  {journal}
  {Phys. Rev. B}\ }\textbf {\bibinfo {volume} {96}},\ \bibinfo {pages} {241401}
  (\bibinfo {year} {2017})}\BibitemShut {NoStop}%
\bibitem [{\citenamefont {Fal’ko}(1992)}]{fal1992cyclotron}%
  \BibitemOpen
  \bibfield  {author} {\bibinfo {author} {\bibfnamefont {V.~I.}\ \bibnamefont
  {Fal’ko}},\ }\href@noop {} {\bibfield  {journal} {\bibinfo  {journal}
  {Physical Review B}\ }\textbf {\bibinfo {volume} {46}},\ \bibinfo {pages}
  {4320} (\bibinfo {year} {1992})}\BibitemShut {NoStop}%
\bibitem [{\citenamefont {Herzog}\ \emph {et~al.}(2017)\citenamefont {Herzog},
  \citenamefont {Hardtdegen}, \citenamefont {Sch\"apers}, \citenamefont
  {Grundler},\ and\ \citenamefont {Wilde}}]{herzog17:103012}%
  \BibitemOpen
  \bibfield  {author} {\bibinfo {author} {\bibfnamefont {F.}~\bibnamefont
  {Herzog}}, \bibinfo {author} {\bibfnamefont {H.}~\bibnamefont {Hardtdegen}},
  \bibinfo {author} {\bibfnamefont {T.}~\bibnamefont {Sch\"apers}}, \bibinfo
  {author} {\bibfnamefont {D.}~\bibnamefont {Grundler}}, \ and\ \bibinfo
  {author} {\bibfnamefont {M.}~\bibnamefont {Wilde}},\ }\href@noop {}
  {\bibfield  {journal} {\bibinfo  {journal} {New Journal of Physics}\ }\textbf
  {\bibinfo {volume} {19}},\ \bibinfo {pages} {103012} (\bibinfo {year}
  {2017})}\BibitemShut {NoStop}%
\bibitem [{\citenamefont {Wilde}\ and\ \citenamefont
  {Grundler}(2013)}]{wilde2013alternative}%
  \BibitemOpen
  \bibfield  {author} {\bibinfo {author} {\bibfnamefont {M.~A.}\ \bibnamefont
  {Wilde}}\ and\ \bibinfo {author} {\bibfnamefont {D.}~\bibnamefont
  {Grundler}},\ }\href@noop {} {\bibfield  {journal} {\bibinfo  {journal} {New
  Journal of Physics}\ }\textbf {\bibinfo {volume} {15}},\ \bibinfo {pages}
  {115013} (\bibinfo {year} {2013})}\BibitemShut {NoStop}%
\bibitem [{\citenamefont {Golub}\ and\ \citenamefont {van
  Loan}(2013)}]{golub:book}%
  \BibitemOpen
  \bibfield  {author} {\bibinfo {author} {\bibfnamefont {G.}~\bibnamefont
  {Golub}}\ and\ \bibinfo {author} {\bibfnamefont {C.~F.}\ \bibnamefont {van
  Loan}},\ }\href@noop {} {\emph {\bibinfo {title} {Matrix Computations}}},\
  \bibinfo {edition} {4th}\ ed.\ (\bibinfo  {publisher} {The Johns Hopkins
  University Press},\ \bibinfo {year} {2013})\BibitemShut {NoStop}%
\bibitem [{Note1()}]{Note1}%
  \BibitemOpen
  \bibinfo {note} {Here we benchmark using a powerful laptop.}\BibitemShut
  {Stop}%
\bibitem [{\citenamefont {Casanova}\ \emph {et~al.}(2010)\citenamefont
  {Casanova}, \citenamefont {Romero}, \citenamefont {Lizuain}, \citenamefont
  {Garc{\'\i}a-Ripoll},\ and\ \citenamefont {Solano}}]{casanova2010deep}%
  \BibitemOpen
  \bibfield  {author} {\bibinfo {author} {\bibfnamefont {J.}~\bibnamefont
  {Casanova}}, \bibinfo {author} {\bibfnamefont {G.}~\bibnamefont {Romero}},
  \bibinfo {author} {\bibfnamefont {I.}~\bibnamefont {Lizuain}}, \bibinfo
  {author} {\bibfnamefont {J.~J.}\ \bibnamefont {Garc{\'\i}a-Ripoll}}, \ and\
  \bibinfo {author} {\bibfnamefont {E.}~\bibnamefont {Solano}},\ }\href@noop {}
  {\bibfield  {journal} {\bibinfo  {journal} {Physical Review Letters}\
  }\textbf {\bibinfo {volume} {105}},\ \bibinfo {pages} {263603} (\bibinfo
  {year} {2010})}\BibitemShut {NoStop}%
\bibitem [{\citenamefont {Braak}(2011)}]{braak11:100401}%
  \BibitemOpen
  \bibfield  {author} {\bibinfo {author} {\bibfnamefont {D.}~\bibnamefont
  {Braak}},\ }\href@noop {} {\bibfield  {journal} {\bibinfo  {journal} {Phys.\
  Rev.\ Lett.}\ }\textbf {\bibinfo {volume} {107}},\ \bibinfo {pages} {100401}
  (\bibinfo {year} {2011})}\BibitemShut {NoStop}%
\bibitem [{\citenamefont {Candido}\ \emph {et~al.}(2023)\citenamefont
  {Candido}, \citenamefont {Erlingsson}, \citenamefont {Gramizadeh},
  \citenamefont {Costa}, \citenamefont {Weigele}, \citenamefont {Zumbühl},\
  and\ \citenamefont {Egues}}]{candido23:cond-mat}%
  \BibitemOpen
  \bibfield  {author} {\bibinfo {author} {\bibfnamefont {D.~R.}\ \bibnamefont
  {Candido}}, \bibinfo {author} {\bibfnamefont {S.~I.}\ \bibnamefont
  {Erlingsson}}, \bibinfo {author} {\bibfnamefont {H.}~\bibnamefont
  {Gramizadeh}}, \bibinfo {author} {\bibfnamefont {J.~V.~I.}\ \bibnamefont
  {Costa}}, \bibinfo {author} {\bibfnamefont {P.~J.}\ \bibnamefont {Weigele}},
  \bibinfo {author} {\bibfnamefont {D.~M.}\ \bibnamefont {Zumbühl}}, \ and\
  \bibinfo {author} {\bibfnamefont {J.~C.}\ \bibnamefont {Egues}},\ }\href@noop
  {} {\enquote {\bibinfo {title} {Quantum oscillations in 2d electron gases
  with spin-orbit and zeeman interactions},}\ } (\bibinfo {year} {2023}),\
  \Eprint {http://arxiv.org/abs/2304.14327} {arXiv:2304.14327
  [cond-mat.mes-hall]} \BibitemShut {NoStop}%
\bibitem [{\citenamefont {Brack}\ and\ \citenamefont
  {Bhaduri}(1997)}]{brack97:book}%
  \BibitemOpen
  \bibfield  {author} {\bibinfo {author} {\bibfnamefont {M.}~\bibnamefont
  {Brack}}\ and\ \bibinfo {author} {\bibfnamefont {R.}~\bibnamefont
  {Bhaduri}},\ }\href@noop {} {\emph {\bibinfo {title} {Semiclassical
  physics}}}\ (\bibinfo  {publisher} {Addison-Wesley Publishing},\ \bibinfo
  {year} {1997})\BibitemShut {NoStop}%
\bibitem [{\citenamefont {Fang}\ and\ \citenamefont
  {Stiles}(1968)}]{fang68:823}%
  \BibitemOpen
  \bibfield  {author} {\bibinfo {author} {\bibfnamefont {F.~F.}\ \bibnamefont
  {Fang}}\ and\ \bibinfo {author} {\bibfnamefont {P.~J.}\ \bibnamefont
  {Stiles}},\ }\href {\doibase 10.1103/PhysRev.174.823} {\bibfield  {journal}
  {\bibinfo  {journal} {Phys. Rev.}\ }\textbf {\bibinfo {volume} {174}},\
  \bibinfo {pages} {823} (\bibinfo {year} {1968})}\BibitemShut {NoStop}%
\bibitem [{\citenamefont {Brosig}\ \emph {et~al.}(2000)\citenamefont {Brosig},
  \citenamefont {Ensslin}, \citenamefont {Jansen}, \citenamefont {Nguyen},
  \citenamefont {Brar}, \citenamefont {Thomas},\ and\ \citenamefont
  {Kroemer}}]{brosig00:13045}%
  \BibitemOpen
  \bibfield  {author} {\bibinfo {author} {\bibfnamefont {S.}~\bibnamefont
  {Brosig}}, \bibinfo {author} {\bibfnamefont {K.}~\bibnamefont {Ensslin}},
  \bibinfo {author} {\bibfnamefont {A.~G.}\ \bibnamefont {Jansen}}, \bibinfo
  {author} {\bibfnamefont {C.}~\bibnamefont {Nguyen}}, \bibinfo {author}
  {\bibfnamefont {B.}~\bibnamefont {Brar}}, \bibinfo {author} {\bibfnamefont
  {M.}~\bibnamefont {Thomas}}, \ and\ \bibinfo {author} {\bibfnamefont
  {H.}~\bibnamefont {Kroemer}},\ }\href {\doibase 10.1103/PhysRevB.61.13045}
  {\bibfield  {journal} {\bibinfo  {journal} {Phys. Rev. B}\ }\textbf {\bibinfo
  {volume} {61}},\ \bibinfo {pages} {13045} (\bibinfo {year}
  {2000})}\BibitemShut {NoStop}%
\bibitem [{\citenamefont {Hatke}\ \emph {et~al.}(2012)\citenamefont {Hatke},
  \citenamefont {Zudov}, \citenamefont {Pfeiffer},\ and\ \citenamefont
  {West}}]{hatke12:241305}%
  \BibitemOpen
  \bibfield  {author} {\bibinfo {author} {\bibfnamefont {A.~T.}\ \bibnamefont
  {Hatke}}, \bibinfo {author} {\bibfnamefont {M.~A.}\ \bibnamefont {Zudov}},
  \bibinfo {author} {\bibfnamefont {L.~N.}\ \bibnamefont {Pfeiffer}}, \ and\
  \bibinfo {author} {\bibfnamefont {K.~W.}\ \bibnamefont {West}},\ }\href
  {\doibase 10.1103/PhysRevB.85.241305} {\bibfield  {journal} {\bibinfo
  {journal} {Phys. Rev. B}\ }\textbf {\bibinfo {volume} {85}},\ \bibinfo
  {pages} {241305} (\bibinfo {year} {2012})}\BibitemShut {NoStop}%
\end{thebibliography}

%

\end{document}